# Developing an Interactive Tutorial on a Quantum Eraser

Emily Marshman and Chandralekha Singh

*Department of Physics and Astronomy, University of Pittsburgh, Pittsburgh, PA, 15260, USA*

**Abstract**: We are developing a quantum interactive learning tutorial (QuILT) on a quantum eraser for students in upper-level quantum mechanics. The QuILT exposes students to contemporary topics in quantum mechanics and uses a guided approach to learning. It adapts existing visualization tools to help students build physical intuition about quantum phenomena and strives to help them develop the ability to apply quantum principles in physical situations. The quantum eraser apparatus in the *gedanken* (thought) experiments and simulations that students learn from in the QuILT uses a Mach-Zehnder Interferometer with single photons. We also discuss findings from a preliminary in-class evaluation.

## INTRODUCTION

Quantum mechanics can be a challenging subject for students partly because it is unintuitive and abstract [1-6]. The Mach-Zehnder Interferometer (MZI) with single photons is an experiment which has been conducted in undergraduate laboratories to illustrate fundamental principles of quantum mechanics [7]. We are developing a quantum interactive learning tutorial (QuILT) on a quantum eraser using *gedanken* (thought) experiments and simulations involving a MZI with single photons. The QuILT focuses on helping students learn topics such as the wave-particle duality of a single photon, interference of a single photon with itself, probabilistic nature of quantum measurements, and collapse of a quantum state upon measurement. Students also learn how photo-detectors (detectors) and optical elements such as beam-splitters and polarizers in the paths of the MZI affect measurement outcomes. In particular, they learn to reason systematically about a quantum eraser setup in which placing a polarizer with specific orientations in a particular location in the MZI setup can result in interference of a single photon with itself due to erasure of "which-path" information (WPI) [7].

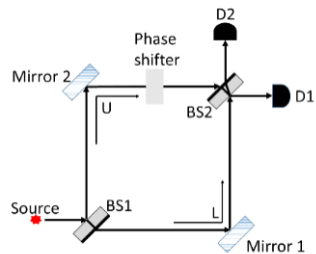

**FIGURE 1**. MZI setup with a phase shifter in the U path

Students are given a schematic diagram of the MZI setup in the QuILT (see the basic setup in Fig. 1). As they work through the QuILT, they are told to make simplifying assumptions about the MZI setup including the following: 1) all optical elements are ideal and polarizers are absorbing, i.e., the photon is either absorbed or transmitted by the polarizer); 2) the non-polarizing beam-splitters (BS1 and BS2) are infinitesimally thin such that there is no phase shift when a single photon propagates through them; 3) the detectors are polarization sensitive, i.e., in a particular basis, the detector can measure the polarization of the photon which is detected; 4) either unpolarized photons (e.g., an equal mixture of horizontally and vertically polarized photons) or monochromatic +45° polarized single photons from the source travel the same distance in vacuum in the upper path (U) and lower path (L) of the MZI; and 5) the initial MZI without the phase shifter is set up such that there is completely constructive interference at detector 1 (D1) and destructive interference at detector 2 (D2).

Using a guided approach to learning, the QuILT helps students reason about how observing interference of a single photon with itself at D1 and D2 can be interpreted in terms of not having WPI about the single photon [7]. WPI is a common terminology associated with these types of experiments popularized by Wheeler [8] and WPI is "known" about the photon if D1 and D2 can only project one component of the photon path state. For example, if BS2 is removed from the setup in Fig. 1, WPI is known for single photons arriving at the detectors because only the component of a photon state along the U path can be projected in D1 and only the component of a photon state along the L path can be projected in D2. When WPI is known, each detector (D1 and D2) has equal probability of clicking. A detector clicks when a photon is detected by it and is absorbed (the state of the single photon collapses, i.e., the single photon state is no longer in a superposition of the U and L path states). When WPI is known for all single photons arriving at the detectors, there is no way to know a priori which detector will click when a photon is sent until the photon state collapses either at D1 or at D2 with equal likelihood. On the other hand, WPI is unknown about single photons arriving at the detectors in the setup shown in Fig. 1 because BS2 mixes the path states of the single photon. Thus, D1 and D2 can project both components of the photon path state and the projection of both components at each detector leads to interference. When WPI is unknown and a large number of single photons are sent through the setup, if a phase shifter is inserted in one of the paths of the MZI (e.g., in the U path in Fig. 1) and its thickness is varied, the probability of photons arriving at D1 and D2 will change with the thickness of the phase shifter due to interference of the components of the single photon state from the U and L paths. When WPI is known, changing the thickness of a phase shifter does not affect the probability of each detector clicking when photons are registered (equal probability for all thicknesses of phase shifter) [7].

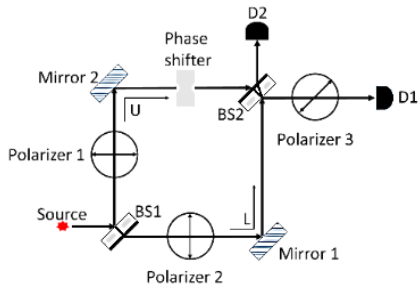

**FIGURE 2.** Quantum Eraser Setup

The guided approach in the QuILT helps students reason about the fact that the setup shown in Fig. 2 is a quantum eraser in which placing polarizer 3 as shown with its orientation (other than vertical or horizontal) erases WPI and results in interference of a single photon with itself at D1 (however, if polarizer 3 is removed from Fig. 2, WPI is known and no interference is observed).

## DEVELOPMENT OF THE QUILT

We developed a preliminary version of the QuILT (which includes a warm-up with background information about the MZI setup and pre-/posttests to be given before and after the QuILT) that uses a guided approach to learning and accounts for common student difficulties discussed later. The QuILT makes use of a computer simulation in which students can manipulate the MZI setup to predict and observe what happens at the detectors for different setups. Different versions of the QuILT were iterated with three physics faculty members several times to ensure that they agreed with the content and wording. We also administered it to several graduate and upper-level undergraduate students to ensure that the guided approach was effective and the questions were unambiguously interpreted. Modifications were made based upon the feedback.

During the development of the QuILT, we investigated the difficulties students have with the relevant concepts in order to effectively address them. We conducted 15 individual semi-structured think-aloud interviews with upper-level undergraduate and graduate students using different versions of an open-ended survey or earlier versions of the QuILT in which students were first asked to think aloud as they answered the questions related to the setup including those with polarizers at various locations (some of the configurations being a quantum eraser) to the best of their ability without being disturbed. Later, we probed students further and asked for clarification of points as needed. Since both undergraduate and graduate students exhibited the same difficulties, we will not distinguish between the two groups further. Some of the common difficulties addressed in the QuILT and summarized below include how a single photon can interfere with itself, how polarizers can act as measurement devices and alter the state of a photon, and how WPI can be erased, e.g., by introducing polarizer 3 in Fig. 2.

**Difficulty with a single polarizer in the U or L path of the MZI:** Interviews suggest that many students had difficulty with how the interference at D1 and D2 in Fig. 1 is affected by placing a single polarizer, e.g., with a vertical polarization axis in the L path of the MZI. In this situation, if the source emits a large number of unpolarized single photons, there are three possible measurement outcomes at the detectors due to the polarizer: 1) the photon is absorbed by the polarizer and it does not reach the detectors D1 or D2 (25% probability); 2) the photon is not absorbed by the vertical polarizer but both the photon path state and polarization state collapse, i.e., the photon has a 25% probability of being in the U path with a horizontal polarization; and 3) the photon is not absorbed by the vertical polarizer and the polarization state of the photon collapses but not the path state, i.e., the photon has a 50% probability of having a vertical polarization and remaining in a superposition of the U and L path states. If a detector registers a photon with a horizontal polarization, WPI is known since the vertical polarizer collapsed the photon with a horizontal polarization to the U path state. However, WPI is unknown if a detector registers a photon with vertical polarization since the vertical polarizer does not collapse the path state of such a photon and this photon displays constructive interference at D1 and destructive interference at D2 in the given setup without the phase shifter. Thus, D1 will register all single photons with a vertical polarization (50% of photons emitted from the source) and 12.5% of the single photons emitted from the source which collapsed to the horizontal polarization state due to the vertical polarizer in the L path. D2 will register only photons with a horizontal polarization (12.5% of the photons emitted from the source). Some students correctly stated that if one polarizer with a vertical polarization axis is placed in the L path, fewer photons would reach the detectors D1 and D2 but they incorrectly claimed that all of the photons that reach the detectors would display interference. For example, one student said: "Some of the photons won't make it to the [detectors]. 75% of the photons display interference because only half of the photons in path [L] will go through." These types of responses indicate that students struggled with the fact that a single polarizer collapses the path state for some of the photons and thus there are some photons that show interference and others that do not show interference at the detectors.

**Difficulty with two orthogonal polarizers placed in the U and L paths of the MZI:** Students often incorrectly claimed that the effect of placing two orthogonal polarizers in the two paths of the MZI is not different from the effect of a single polarizer in one path except that fewer photons would reach the detectors. In this case, the two orthogonal polarizers collapse the photon path state to either the U or L path state. WPI is known about all photons arriving at the detectors, interference is destroyed, and the detectors register photons with equal probability. Many students stated that "fewer photons would reach the detectors" but that interference would still be displayed. For example, one student stated: "50% of the photons emitted by the source display interference because we don't measure anything until the photons hit the [detector] so their state vector doesn't collapse until then." These students struggled with the fact that two orthogonal polarizers placed in the two paths of the MZI correspond to a measurement of photon polarization. Either the photon gets absorbed by the polarizer or the photon with a vertical polarization that reaches D1 or D2 came only from the MZI path with the vertical polarizer and the photon with a horizontal polarization that reaches D1 or D2 came only from the path with the horizontal polarizer. Thus, WPI is known about all photons that reach D1 and D2. Regarding Fig. 2 without polarizer 3, students had difficulty with the fact that once the photon reaches the polarizers, the measurement of polarization collapses the state of the photon such that if a detector registers a photon with a horizontal polarization, it must have come from the U path and if a detector registers a photon with a vertical polarization, it must have come from the L path. WPI is known for all photons and interference is destroyed.

**Difficulty with a quantum eraser setup**: In contrast to the MZI setup with two orthogonal polarizers in which WPI is known about all photons arriving at the detectors regardless of whether the photons are unpolarized or +45° polarized, the addition of the third polarizer (see Fig. 2) causes both components of the photon path state to be projected into detector D1 for a +45° polarized photon, erasing WPI about the photons arriving at D1. Therefore, if a phase shifter is inserted in one of the paths of the MZI and its thickness is gradually changed, the interference displayed at D1 will change. Some students incorrectly claimed that the quantum eraser setup is not different from the setup in which two orthogonal polarizers are placed in the U and L paths except fewer photons would reach D1 because some will be absorbed by polarizer 3 (see Fig. 2). Moreover, many students could not articulate why the quantum eraser setup shows interference effects at D1 and the setup with two orthogonal polarizers placed in the U and L paths of the MZI does not show interference. For example, one student correctly said that in the quantum eraser setup, "25% of the photons will display interference because only half of the photons going through BS2 will make it through." However, he did not differentiate between the quantum eraser setup and the setup without polarizer 3, incorrectly claiming there would be interference at D1 in both cases. Some students stated that none of the photons would display interference in Fig. 2, e.g., "0% display interference, they are all independent photons." These types of responses indicate that some students have difficulty with the role of the polarizer 3.

**Assuming that *any* photon reaching the detectors displays interference, regardless of the polarizers in the setup:** Some students assumed that any photon reaching the detectors would display interference, regardless of the polarizers in the setup. For example, one student incorrectly stated that "all of the photons display interference since…every photon splits between both paths." He used this incorrect reasoning to explain that all of the photons that reach the detectors would display interference in all setups (with one polarizer, two orthogonal polarizers, and the quantum eraser). Other students who correctly determined the percentage of photons that would pass through the polarizers and arrive at D1 or D2 incorrectly stated that any photon reaching the detectors would display interference. For example, regarding the setup with a single polarizer with a vertical polarization axis in the L path, one student incorrectly claimed that "75% of the photons display interference because 50% of the initial photons take path U and make it to the [detector] and 50% of the 50% taking path L make it so 50% + 25% = 75%." These types of responses indicate that students had difficulty with how one or more polarizers placed in the paths of the MZI change the photon state and may provide WPI for some or all of the photons (destroying interference for some or all photons at the detectors).

## PRELIMINARY EVALUATION

Once we determined that the QuILT was effective in individual administration, it was given to 18 upper-level undergraduates in a first-semester quantum mechanics course and 18 first-year graduate students. Students were first given a pretest. They then worked through the QuILT in class and were asked to complete whatever they could not finish in class as homework. Then, a posttest was administered which had analogous questions as the pretest except that the orientations of the polarizers differed (e.g., instead of vertical and horizontal polarizers in the two paths with a source emitting +45° polarized single photons, the posttest had +45° and -45° polarizers in the two paths with a source emitting vertically polarized single photons).

Table 1 shows the common student difficulties and percentages exhibiting those difficulties on the pretest and posttest. Table 2 shows the average percentage scores on the pre-/posttest questions. Part (a) of each question asks students to compare two different MZI setups with polarizers and describe how they are different, e.g., "You insert a polarizer with a vertical polarization axis in the U path of the MZI. Describe what you would observe at D1 and D2 and how this situation will differ from the case in which there is no polarizer in path U." Part (b) asks for the percentage of photons that display interference. The average normalized gain from pretest to posttest was 0.72 [9].

**TABLE 1**. Common difficulties and percentages of 36 students displaying them on the pre-/posttest questions.

| | |
|---|---|
| Q1 One polarizer in the path of the MZI will not change the interference | 42/11 |
| Q2 A MZI setup with two orthogonal polarizers placed in the two paths (one in each path) is not different from the setup with one polarizer except fewer photons reach the detector and interference is displayed regardless of the polarizer setup | 31/8 |
| Q3 The quantum eraser setup is not different from placing two orthogonal polarizers in the two paths of the MZI except fewer photons reach the detectors | 28/8 |

**TABLE 2**. Average percentage scores on the pretest and posttest questions for 36 students.

| Question | 1a/1b | 2a/2b | 3a/3b |
|---|---|---|---|
| Pretest | 22/17 | 54/53 | 39/36 |
| Posttest | 78/72 | 83/83 | 89/89 |

Question 1 on the pre-/posttest assessed student understanding of the effect of one polarizer placed in one of the paths of the MZI. Students were asked to explain how inserting a polarizer with a vertical polarization axis in the U path of the MZI would affect what happens at the detectors compared to the original MZI setup in which there is no polarizer (Fig. 1) and they were asked to write down the percentage of photons displaying interference at the detectors. In the pretest, 42% of the students correctly claimed that the single polarizer would absorb some of the photons and thus fewer photons would reach the detectors, but they incorrectly claimed that all of the photons reaching the detector would display interference. After working on the QuILT, the difficulty with how one polarizer will affect the interference was reduced (see Table 1).

Question 2 on the pre-/posttest assessed student understanding of the effect of placing two orthogonal polarizers in the two paths of the MZI. Students were asked to describe how this situation is different from the case in which there was only one polarizer present and what percentage of photons would display interference. After working through the QuILT, the difficulty with how two orthogonal polarizers affect the interference at the detectors was reduced (see Table 1).

Question 3 on the pre-/posttest assessed student understanding of a quantum eraser (see Fig. 2). The addition of the third polarizer causes both components of the photon path state to be projected at the detector D1, erasing WPI about the photons arriving at D1. As the thickness of the phase shifter is varied, the interference displayed at D1 will change (unlike the setup without polarizer 3). In the pretest, 28% of students incorrectly claimed that the quantum eraser setup is not different from the setup with two orthogonal polarizers in the paths of the MZI or that fewer photons would reach the detectors but otherwise the setups are the same. After working through the QuILT, this difficulty was reduced (see Table 1).

The difficulty related to incorrectly assuming that *any* photon reaching the detectors displays interference regardless of the polarizers in the setup was displayed in questions 1, 2, and 3. After working through the QuILT, this difficulty was reduced (see Tables 1 and 2).

## SUMMARY AND FUTURE PLANS

The quantum eraser QuILT uses a MZI experiment with single photons to help students learn how polarizers affect the interference of a single photon with itself in an exciting context. By taking into account students' prior knowledge and difficulties, the QuILT helps students learn how interference at the detectors in the MZI setup can be restored by introducing a third polarizer with a certain orientation between BS2 and a detector. Many students stated that it was one of their favorite QuILTs and they were excited to be introduced to contemporary topics in quantum mechanics. We are developing a related QuILT using a product space of the two state systems for the photon path and polarization states to help students connect qualitative understanding of a quantum eraser with mathematical formalism.

## ACKNOWLEDGEMENTS

We thank the National Science Foundation for awards PHY-0968891 and PHY-1202909.